# Continuous, Dynamic and Comprehensive Article-Level Evaluation of Scientific Literature

Xianwen Wang*, Zhichao Fang and Yang Yang

WISE Lab, School of Public Administration & Law, Dalian University of Technology, Dalian 116085, China.

* Corresponding author.
Email address: xianwenwang@dlut.edu.cn

**Interactive visualization of this research:** http://xianwenwang.com/research/ale/

**Abstract:** It is time to make changes to the current research evaluation system, which is built on the journal selection. In this study, we propose the idea of continuous, dynamic and comprehensive article-level-evaluation based on article-level-metrics. Different kinds of metrics are integrated into a comprehensive indicator, which could quantify both the academic and societal impact of the article. At different phases after the publication, the weights of different metrics are dynamically adjusted to mediate the long term and short term impact of the paper. Using the sample data, we make empirical study of the article-level-evaluation method.

**Keywords:** *Article-level-metrics, Article-level-evaluation, Altmetrics, Essential Science Indicators, Nature Index, Web of Science*

## Introduction

For decades, citation has been regarded as the sole indicator to evaluate the impact of a paper, the paper which is cited more frequently means the research results gained more recognition. However, citations need a long time (often over two years) to accumulate. In many situations, e.g., funding decisions, hiring tenure and promotion, people need to make evaluations for newly published papers. Alternatively, some people begin to use journal based metrics, e.g., Journal Impact Factor, as an alternative way to quantify the qualities of individual research articles(Alberts, 2013). There are many debates about the abuse of Impact Factor (Bordons, Fernández, & Gomez, 2002; Garfield, 2006; Opthof, 1997; PLoS_Medicine_Editors, 2006; Seglen, 1997), applying Journal Impact Factor to assess the research excellence is not the right way. In addition, only tracking citation metrics could not tell the whole story about the influence of a paper. Besides citation, the impact of scientific papers could be reflected with article usage (browser views and pdf downloads), captures (bookmarks and readership), online mentions (blog posts, social media discussions and news reports)(Priem, Taraborelli, Groth, & Neylon, 2010). Therein, the idea of altmetrics comes into being. Different from citation, which puts particular emphasis on describing the academic impact of articles, altmetrics is based on data gathered from social media platforms and focuses on the societal impact. Compared with the long time for papers to reach their citation peaks, it takes a short period for newly published articles to peak for altmetric scores. In summary, citation is an indicator to measure the long term academic impact, when the indicator of altmetrics reflects short term societal impact. Neither citations nor altmetrics individually could fully indicate the complete impact of a paper, we cannot accurately conjecture the results of one metric by the results of another.

It is necessary to find a new way to quantify both the academic and societal impact together, and mediate the long term and short term impact of the paper. Some publishers have already listed the different types of metrics for an individual article, e.g., PLOS, when some altmetrics tools and services are also available, e.g., Impact Story, Altmetric.com, Plum Analytics, etc. Although altmetric score from altmetric.com is a weighted count that integrates different online mentions of the paper. If we go further on this way, taking all available metrics (e.g., citation, usage, online attention, etc.) into consideration to design a comprehensive metric, which could be used to evaluate the complete impacts of articles.

## The absence of evaluating data source

According to the official statement of Web of Science, it is designed for researchers to "find high-impact article". Nowadays, with the absence of specialized evaluating data source, Web of Science has been adopted by many scientometrics researchers and institutions as the primary data source of article evaluation. In some countries, e.g., China, articles indexed in Science Citation Index/Social Science Citation Index or not is an important criterion to judge the quality of the research.

However, applying Web of Science to assess the research performance and research excellence is not a good choice. Web of Science is designed and created on the basis of journal selection, it collectively index journals cover-to-cover. However, articles published in the same journal, the same issue, have totally different impacts. Even for those high impact factor journals, there are many articles have few citations.

We check the articles published in 2000 and indexed in Science Citation Index Expanded, as Table 1 shows. For example, 2901 of the total 13660 articles in Chemical Engineering have never been cited. For the area of Condensed Matter Physics, the zero-citation percentage is 10.91%, for the area of Biochemistry, Molecular Biology, the zero-citation percentage is 3.23%.

Table 1 Number of Zero-citation articles in 2000 indexed in Science Citation Index Expanded

|  | Total | Zero-citation | Percentage |
| --- | --- | --- | --- |
| Engineering, Chemical | 13660 | 2901 | 21.24% |
| Physics, Condensed Matter | 21974 | 2397 | 10.91% |
| Biochemistry, Molecular Biology | 42710 | 1380 | 3.23% |

There are also some publishers regard Web of Science as a profit-making tool. For example, *Academic Journals* charges a US$550-$750 manuscript handling fee from the author for each accepted article (http://www.harzing.com/esi_highcite.htm). Among which, several ISI-listed journals publish more than 1,000 articles per year, e.g., in 2007, *African Journal of Business Management* only published 28 articles, in 2010, it published 446, when in 2011, as many as 1350 articles were published by this single journal. Thomson Reuters has the mechanism to review the exiting journal coverage constantly, some journals that have become less useful would be deleted. However, this kind of mechanism does not apply to the articles, even some journals are deleted from the coverage, numerous low-quality papers published by these journals are still indexed in Web of Science.

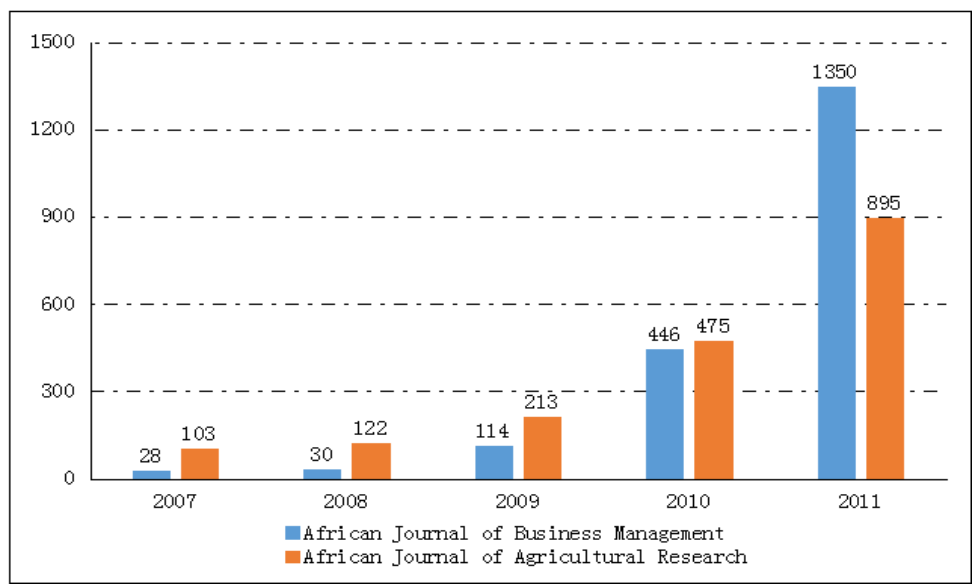

Figure 1 Rapid growth of yearly indexed articles of two journals

With the same idea of Web of Science, Nature Publishing Group (NPG) introduced the Nature Index in November 2014, which is "a database of author affiliation information collated from research articles published in an independently selected group of 68 high-quality science journals". The 68 journals are selected by a group of professors and validated by 2,800 responses to a large-scale survey, when these 68 journals account for approximate 30% of total citations to natural science journals. (http://www.nature.com/press_releases/nature-index.html)

Based on journal article publication counts and citation data from Thomson Scientific databases (mainly from Web of Science), ISI/Thomson (now Thomson Reuters) proposed Essential Science Indicators (ESI), which is an in-depth analytical tool and also a database where citations are analyzed, so that scientists, journals, institutions, and countries can be ranked and compared, for example, most cited scientists rankings, institutions rankings and countries rankings. Ranking in ESI is made by the citations, it has nothing to do with the Impact Factors of journals, which means that whichever journal the paper is published in, citations is the only factor to be taken into account.

Although ESI set a relatively low selection criterion for newly published papers (http://www.in-cites.com/thresholds-highly-cited.html), using cited times to evaluate is not a good choice.

Compared to 8670 journals covered by Science Citation Index Expanded, the journals selected by Nature Index is so much less, which makes Nature Index become an elite database. The aim of Nature Index is "intended to be one of a number of metrics to assess research excellence and institutional performance" (http://www.natureindex.com/faq). However, we think journal-based database is not appropriate for research evaluation, including research excellence and institutional performance, which should be on the basis of article-level metrics. Because of the great influence of Nature Publishing Group, the Nature Index will definitely make great changes to the academia and research evaluation system.

It is necessary to make changes to the current evaluating way of scientific literature. In this research, our purpose is to design a new method, through which the continuous, dynamic and comprehensive evaluation of scientific literature could be made. This new method will be valuable to the research community. With this evaluating method and system, we could make a better evaluation of articles, scientists, journals, institutions, and even countries.

## Design a new evaluation way

### *Considering both academic and societal impact of a paper*

As mentioned above, the impact of a paper could be measured by citation, article usage and online mentions, etc., as Table 2 shows.

Table 2 Types and metrics of the impact of a paper

| Type | Metric |
| --- | --- |
| Article usage | browser views (abstract, full-text), pdf downloads |
| Captures | bookmarks (CiteUlike), readers (Mendeley) |
| Online mentions | blog posts, news reports, likes (Facebook), shares (Facebook), Tweets, +1 (Google plus) |
| Citations | citations |

The Issue 6, Volume 8 of PLOS Computational Biology is selected as our research object. It was published in June 2012, and includes 46 research articles.

In November 2012, PLOS began to provide a regular report covering a wide range of article-level-metrics covering all of its journals via the platform http://article-level-metrics.plos.org/. In this research, the cumulative article-level-metrics data for the entire PLOS corpus are harvested from the PLOS ALM platform. From October 2012 to October 2014, PLOS has provided the ALM reports for 8 times, when the provided date are Oct. 10, 2012, Dec. 12, 2012, Jan. 8,

2013, Apr. 11, 2013, May. 20, 2013, Aug. 27, 2013, Mar. 10, 2014 and Oct. 1, 2014. Factor analysis is employed to study the metrics data of the 46 articles, Table 3 shows the results of the data extracted from the ALM report of Oct. 2014.

7 metrics data of Oct. 10, 2012 are factor analyzed by using principal component analysis with Varimax (orthogonal) rotation. The analysis yields two factors explaining a total of 73.709% of the variance for the entire set of variables. Factor 1 is labeled academic impact to the high loadings by the following items: CiteUlike bookmarks, Mendeley readership, PDF downloads and Scopus citations. This first factor explained 48.691% of the variance. The second factor derived is labeled societal impact. This factor is labeled as such due to the high loadings by the two indicators of Facebook and Twitter. The variance explained by this factor is 25.018%. For the indicator of HTML views, the both factor loadings are greater than 0.65, which means that browser HTML views has both academic and societal impact.

Table 3 Rotated Component Matrix

|  | Factor 1: Academic impact | Factor 2: Societal impact |
|---|---|---|
| CiteUlike | 0.775 |  |
| Mendeley | 0.856 |  |
| HTML views | 0.692 | 0.672 |
| PDF downloads | 0.917 |  |
| Scopus | 0.751 |  |
| Facebook |  | 0.745 |
| Twitter |  | 0.709 |

*Note*. Factor loadings < .5 are suppressed

The Altmetric score is a quantitative measure of the attention that a scholarly article has received. It is a weighted count of the different online platform sources (newspaper stories, tweets, blog posts, comments) that mention the paper. Downloads, citations and reader counts from Mendeley or CiteULike are not used in the score calculation. So, Altmetric score could be regarded as a comprehensive indicator that measures the societal impact of paper partially.

*Dual function of societal impact*

The value of societal metrics is not only reflected by the social effects of the diffusing of the knowledge embodied in the literature, but also reflected by the possible additional academic impact caused by social online attention.

Social media make the research achievements and scientific discoveries spread to the general public, which is just the goal of scientific researches. From the other hand, wide spreading of scientific literature could lead to more scholarly citations. The mechanism from online attention to citation is very complicated, but social attention do have the potentiality to contribute some extra citations to a paper (Wang, Liu, Fang, & Mao, 2014; Wang, Mao, Zhang, & Liu, 2013).

*Dynamic patterns of article-level metrics*

For the 46 selected articles published in June 2012, we sum the metrics data at the 8 time periods separately, as Figure 2 shows. Different metrics show different dynamic evolution patterns. In October 2012, when the articles had been published for about 4 months, there is few citations. The curve of citations begins a sharp rise at the phase of May 2013, one year after the publication. However, for the Facebook and Twitter data, the two curves have almost reached their summits at the very first phase. During the next periods, there is little increase for the Facebook and Twitter data. And for the views data, which is placed on the secondary Y axis in Figure 2, the situation is somehow between the citations and Facebook/Twitter. At the first phase, there is considerable data. During the following 7 periods, there is a steady growth trend for the curve of views.

Dynamic patterns for the different metrics are distinct. Social attention comes to go, citation takes a long time to know, when article view also comes fast but keeps a steady growth.

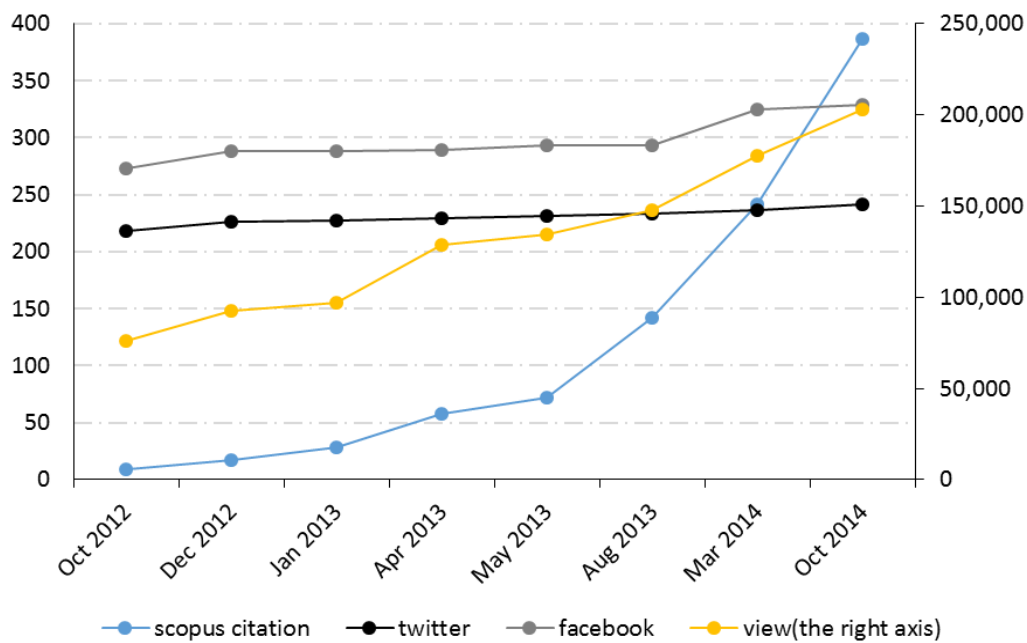

Figure 2 Temporal trend of different metrics of 46 articles published in June 2012

*Article-level evaluation based on Article-level-metrics*

In the era of print, the article could not be separated from the whole issue. For example, libraries could provide the borrowing statistical data, however, it's difficult to know which single article or articles readers are interested in. In the digital era, the situation has been changed greatly. Metrics data for each article are easy to know, including the views, downloads, altmetric score and citations. Of course, some data are easy for publishers to know but not released to public. As early in March 2009, PLOS inaugurated a program to provide "article-level metrics" on a article across all PLOS journals. The metrics data include five main categories, which are Viewed, Cited, Saved, Discussed and Recommended. Following PLOS, more and more publishers began to provide detailed article-level metrics data for readers and researchers. For example, in October 2012, Nature began to provide a real-time online count of article-level metrics for its published research papers, including citation data, news mentions, blog posts and details of sharing through social networks, such as Facebook and Twitter (http://www.nature.com/news/nature-metrics-1.11681). In 2014, the article-level metrics data are also available for PNAS and Science.

The growing article-level metrics dataset provides us with the possibility to design a new evaluating way to make article-level evaluation.

### Problems need to be solved
**Too many indicators**

Citation has been regarded as the single indicator for the past tens of years, nowadays there are much more indicators which are worth being considered, including article views, bookmarks and readership, online discussion, news reports and citations, etc. So many indicators mean a lot of dimensions of the impact, different papers may have different values for the indicators, for example, paper A has been downloaded many times but retweeted few times, when paper B may has opposite situation, so it is very difficult to compare the impact of these two articles, especially when these articles are newly published.

Could these so many indicators be synthesized to one single comprehensive indicator, which could reflect the most of information of the original data and make the papers in diverse situations comparable?

**Dynamic adjustment of the results**

At different phases after publication, the same indicator may have different effects on the impact of the paper. For the newly published articles, because the citations are generally low, it is difficult to judge the qualities and compare the new articles. At the early phase, it is a better choice to use article usage data, online mention data to make evaluation of the newly published articles. As time goes by, the evaluation is gradually dominated by citation metrics, which means that citation would play the most important role in the evaluation when the article has been published for a relatively long time.

To solve these two problems, we propose the idea of designing a comprehensive indicator to reflect all the impacts of an article. The weights of the indicators at different phases should be adjusted dynamically due to the change of relative importance of metrics, just like Table 4 shows.

Table 4 Relative importance of metrics at different phases

| Phase | Relative importance | Selection standard |
|---|---|---|
| 1 (0-6 months) | PDF downloads > HTML views > Twitter > Facebook > Mendeley > CiteUlike > Citation | Top 80% of all articles of same month and subject |
| 2 (6 months-2 years) | PDF downloads > HTML views > Mendeley > CiteUlike > Citation > Twitter > Facebook | Top 70% of all articles of same month and subject |
| 3 (2 -5 years) | Citation > Mendeley > CiteUlike > PDF downloads > HTML views > Twitter > Facebook | Top 50% of all articles of same year and subject |
| 4 (5 years-) | Citation > Mendeley > CiteUlike > PDF downloads > HTML views > Twitter > Facebook | Top 30% of all articles of same year and subject |

To integrate different metrics into a comprehensive indicator, the first problem needs to be solved is weighting. Here we use Analytic hierarchy process (AHP) to calculate the weights of different metrics. The AHP methodology was developed by Thomas L. Saaty in the 1970s (Saaty, 1980). It allows users to assess the relative weight of multiple criteria in an intuitive manner, so it has both advantages of quantitative criteria and qualitative judgment provided by the users. Using pairwise comparisons (X is more important than Y), the relative importance (priority) of one criterion over another can be expressed. To calculate the weights for the different criteria, a pairwise comparison matrix needs to be created. The matrix is a matrix A, where m is the number of evaluation criteria considered, denotes the entry in the $i$th row and the $j$th column of matrix. Each entry of the matrix represents the importance of the $i$th criterion relative to the $j$th criterion. If the cell value in the entry is greater than 1, then the $i$th criterion is more important than the $j$th criterion, and vice versa. If two criteria have the same importance, then the cell value in the entry is 1. The relative importance between two criteria is measured according to a numerical scale from 1 to 9 or 1/9 to 1.

According to the definition of relative importance of different metrics, we need to construct different pairwise comparison matrixes at different phases. The pairwise comparison matrix at phase 1 is shown in Table 5. The higher the weight is, the more important the corresponding criterion becomes, which is represented by the cell value in the matrix. For example, the values in the cells where the row of CiteUlike, the column of HTML views and PDF downloads intersect are less than 1, moreover, the ratio of CiteUlike and PDF downloads is less than the ratio of CiteUlike and HTML views, it means that at phase 1, CiteUlike is less important than HTML views, and much less important than PDF downloads.

Table 5 Pairwise Comparison Matrix at phase 1

|  | CiteUlike | Mendeley | HTML views | PDF downloads | Citation | Facebook | Twitter |
|---|---|---|---|---|---|---|---|
| CiteUlike | 1 | 1 | 1/4 | 1/6 | 4 | 1/4 | 1/6 |
| Mendeley |  | 1 | 1/4 | 1/6 | 4 | 1/4 | 1/6 |
| HTML views |  |  | 1 | 1/4 | 6 | 3 | 2 |
| PDF downloads |  |  |  | 1 | 9 | 4 | 3 |
| Citation |  |  |  |  | 1 | 1/4 | 1/7 |
| Facebook |  |  |  |  |  | 1 | 1/2 |
| Twitter |  |  |  |  |  |  | 1 |

At phase 4, there is much change in the relative importance of the metrics, as Table 6 shows. CiteUlike and Mendeley become more important than HTML views, so the cell values get greater than 1 . At this phase, citation is the most important criterion.

Table 6 Pairwise Comparison Matrix at phase 4

|  | CiteUlike | Mendeley | HTML views | PDF downloads | Citation | Facebook | Twitter |
|---|---|---|---|---|---|---|---|
| CiteUlike | 1 | 1 | 3 | 2 | 1/7 | 3 | 2 |
| Mendeley |  | 1 | 3 | 2 | 1/7 | 3 | 2 |
| HTML views |  |  | 1 | 1/4 | 1/9 | 1 | 1 |
| PDF downloads |  |  |  | 1 | 1/6 | 1 | 1 |
| Citation |  |  |  |  | 1 | 4 | 3 |
| Facebook |  |  |  |  |  | 1 | 1/2 |
| Twitter |  |  |  |  |  |  | 1 |

In this study, the weights and CI values of AHP models are calculated by a CGI system (http://www.isc.senshu-u.ac.jp/~thc0456/EAHP/AHPweb.html). The results are shown in Table 7.

Table 7 Weights of AHP models at different phases

|  | CiteUlike | Mendeley | HTML views | PDF downloads | Citation | Facebook | Twitter |
|---|---|---|---|---|---|---|---|
| Phase 1 | 0.0477 | 0.0477 | 0.1996 | 0.3901 | 0.0234 | 0.1109 | 0.1806 |
| Phase 2 | 0.1723 | 0.1723 | 0.1182 | 0.2108 | 0.1321 | 0.0828 | 0.1116 |
| Phase 3 | 0.1514 | 0.1514 | 0.0481 | 0.0921 | 0.3979 | 0.0644 | 0.0947 |
| Phase 4 | 0.1269 | 0.1269 | 0.0455 | 0.0809 | 0.4819 | 0.0570 | 0.0810 |

In Figure 3, we show the change of the weights of metrics. At Phase 1 and 2, the metric of PDF downloads has the greatest weight. From Phase 1 to 4, the curve of PDF downloads shows a downward trend, when the weight of citation is upward.

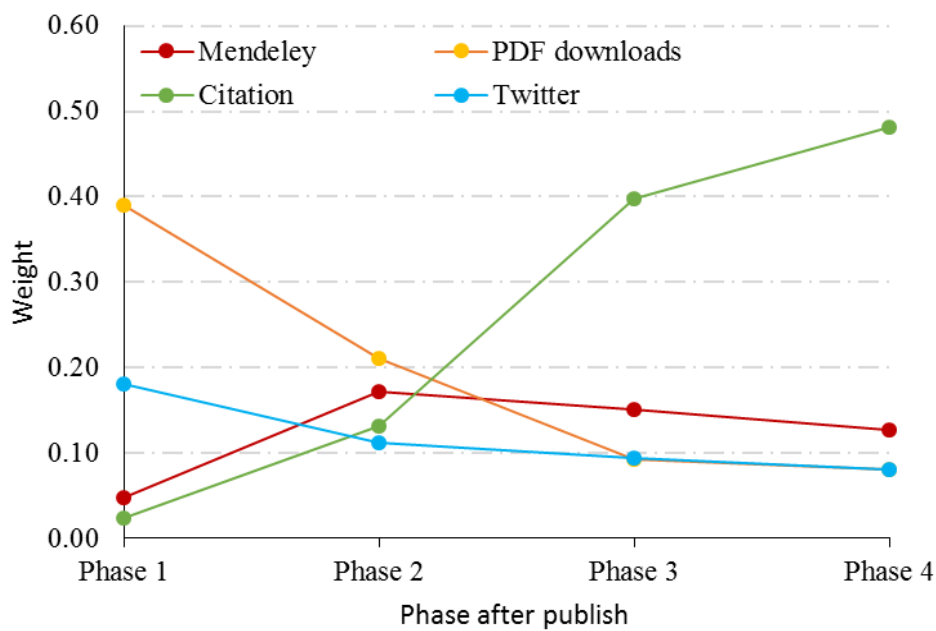

Figure 3 The change of the weights of different metrics

## Empirical Study

The weights in Table 7 are applied to calculate the comprehensive scores of the metrics data of the 46 articles. Metrics data of Oct. 10, 2012 is calculated with the weights of phase 1, when weights of phase 2 and 3 are used for metrics data of Aug. 27, 2013 and Oct. 1, 2014 separately.

All the original metrics data are normalized to the range of 0-1. The normalized value of $e_i$ for variable E in the $i$th row is calculated as:

$$Normalized\ (e_i) = \frac{e_i - E_{min}}{E_{max} - E_{min}}$$

where

$E_{min}$ = the minimum value for variable E

$E_{max}$ = the maximum value for variable E

In Table 8, the values of 7 metrics are original data, when the scores are calculated with the normalized data instead of the original metrics data.

Table 8 lists the top 11 (top 25% of 46) articles of each phase. At phase 1, when the 46 articles had been published for 4 months, article 10.1371/journal.pcbi.1002358 has 16 CiteUlike bookmarks, 81 mendeley readers, 5060 HTML views, 1733 PDF downloads and 3 Scopus citations, etc., when the comprehensive score of this article is 0.7906, ranks top 1. At phase 2, the values of the metrics of Mendeley, HTML views, PDF downloads and Scopus citations have risen sharply, but not for the metrics of Facebook and Twitter, when the score is 0.8579 and still ranks top 1. From phase 1 to 2 and 3, there is much change for the top 11 articles. The ranks of some articles rise, when others may fall. For example, article 10.1371/journal.pcbi.1002538 ranks 6[th] at phase 1, downs to 9 at phase 3, and is disappeared from the top 11 at phase 3; article 10.1371/journal.pcbi.1002531 ranks 11 at phase 1, and rises to top 4 at phase 3.

Table 8 Top 25% articles with greatest score at 3 phases

| phase | rank | doi | citeulike | mendeley | html | pdf | citation | facebook | twitter | score |
|---|---|---|---|---|---|---|---|---|---|---|
| 1 | 1 | 1002358 | 16 | 81 | 5060 | 1733 | 3 | 8 | 12 | 0.7906 |
|  | 2 | 1002543 | 14 | 0 | 4041 | 871 | 0 | 2 | 31 | 0.5653 |
|  | 3 | 1002590 | 0 | 18 | 4302 | 469 | 0 | 73 | 11 | 0.4413 |
|  | 4 | 1002561 | 3 | 37 | 3579 | 721 | 0 | 0 | 9 | 0.3671 |
|  | 5 | 1002519 | 3 | 17 | 2516 | 648 | 0 | 0 | 13 | 0.3146 |
|  | 6 | 1002538 | 3 | 6 | 1777 | 394 | 0 | 22 | 15 | 0.2603 |
|  | 7 | 1002541 | 13 | 24 | 1794 | 354 | 0 | 3 | 12 | 0.2456 |
|  | 8 | 1002527 | 3 | 12 | 1818 | 373 | 0 | 6 | 14 | 0.2305 |
|  | 9 | 1002572 | 6 | 18 | 2045 | 489 | 0 | 0 | 6 | 0.2248 |
|  | 10 | 1002588 | 0 | 13 | 1809 | 454 | 1 | 0 | 7 | 0.1989 |
|  | 11 | 1002531 | 4 | 20 | 1519 | 522 | 1 | 2 | 1 | 0.1865 |
| 2 | 1 | 1002358 | 16 | 170 | 11720 | 3236 | 30 | 7 | 14 | 0.8579 |
|  | 2 | 1002543 | 16 | 72 | 5389 | 1103 | 1 | 2 | 34 | 0.4739 |
|  | 3 | 1002561 | 3 | 79 | 9669 | 1242 | 5 | 2 | 11 | 0.3408 |
|  | 4 | 1002541 | 15 | 57 | 3609 | 665 | 3 | 4 | 13 | 0.3395 |
|  | 5 | 1002590 | 1 | 36 | 6024 | 627 | 1 | 91 | 13 | 0.2622 |
|  | 6 | 1002531 | 8 | 39 | 3389 | 912 | 11 | 3 | 1 | 0.2552 |
|  | 7 | 1002519 | 3 | 39 | 5515 | 1262 | 1 | 0 | 13 | 0.2419 |
|  | 8 | 1002572 | 6 | 44 | 3273 | 754 | 2 | 0 | 6 | 0.2006 |
|  | 9 | 1002538 | 3 | 14 | 3155 | 668 | 4 | 22 | 15 | 0.1889 |
|  | 10 | 1002577 | 2 | 25 | 5063 | 1141 | 2 | 0 | 5 | 0.1816 |
|  | 11 | 1002527 | 3 | 21 | 3266 | 638 | 1 | 6 | 14 | 0.1641 |
| 3 | 1 | 1002358 | 18 | 324 | 19909 | 4651 | 73 | 23 | 14 | 0.8942 |
|  | 2 | 1002543 | 16 | 95 | 6071 | 1241 | 1 | 2 | 36 | 0.3113 |
|  | 3 | 1002541 | 16 | 91 | 4896 | 824 | 11 | 4 | 13 | 0.2931 |
|  | 4 | 1002531 | 9 | 77 | 5670 | 1229 | 26 | 3 | 1 | 0.2874 |
|  | 5 | 1002561 | 4 | 121 | 11231 | 1577 | 21 | 2 | 11 | 0.2866 |
|  | 6 | 1002588 | 0 | 56 | 6112 | 1314 | 19 | 3 | 8 | 0.1849 |
|  | 7 | 1002572 | 9 | 62 | 3803 | 910 | 6 | 0 | 6 | 0.1707 |
|  | 8 | 1002519 | 3 | 69 | 8233 | 1653 | 6 | 0 | 13 | 0.1692 |
|  | 9 | 1002590 | 1 | 42 | 7101 | 904 | 3 | 90 | 13 | 0.1690 |
|  | 10 | 1002555 | 3 | 31 | 5048 | 701 | 13 | 22 | 4 | 0.1531 |
|  | 11 | 1002562 | 7 | 58 | 2840 | 529 | 10 | 0 | 0 | 0.1476 |

Note: (1) Because of the limited layout space, the first half of the doi is omitted. For example, for the doi 10.1371/journal.pcbi.1002358, we only keep 1002358 in Table 8.

(2) Detailed information of Table 8 is available at http://xianwenwang.com/research/ale

The dynamic changes of the scores and rankings of the 46 articles from phase 1 to 3 are shown in Figure 4. The DOIs of 46 articles are listed on the leftmost column, and ranked according to the scores at phase 1. The position of article at the certain phase is decided by the ranking of score at that phase. 46 articles could be only compared at the same phase. Articles at different phases, and even the same article at different phases are not comparable. As shown in Figure 4, if the rank of an article from phase 1 to 3 shows an upward trend, it is displayed with a red curve, there are 20 papers with red curves. We use green curve to represent the downward trend, there are also 20 papers with green curves. Otherwise, if the rank of the article has not changed, the color of the curve is yellow, there are 6 yellow curves. In Figure 4, one red curve with dramatic upward trend is highlighted, indicating that the performance of this paper is rising. The doi of this article is 10.1371/journal.pcbi.1002552, it only ranks 37 at phase 1, rises to 28 at phase 2 and continue to rise to 13 at phase 3.

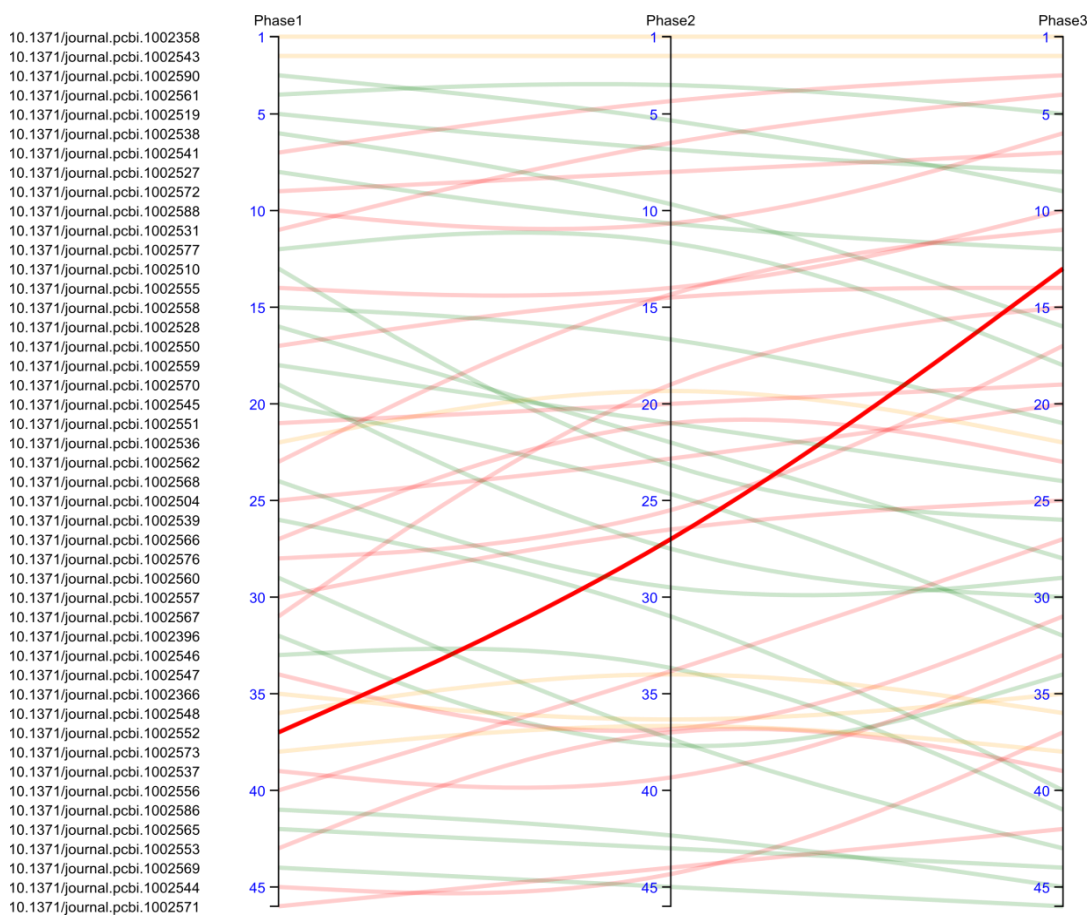

Figure 4 Dynamic changes according to the ranking at different phases

Detailed information of Table 8 is available at

http://xianwenwang.com/research/ale/dynamic.html

## Discussion

In the 1950s, people read papers from printed journals. A group of articles are bundled together to form an issue of journal, it is difficult to separate single article from the whole issue, which is the carrier of articles. For example, if we want to know which paper the readers are interested in when they borrow the journal from the library, that seems to be an extremely difficult task. At that time, journal evaluation is the most important and basic issue. SCI is designed on the basis of core journals selection, specialized indicators and tools are proposed to evaluate journals, e.g., Impact Factor and Journal Citation Reports.

Compared to fifty years ago, scholarly communicating ways have changed a lot. With the advent and fast development of computers, internet and digital libraries, the transformation from print to electronic publishing is accelerating, just as the digital music revolution set music free from the carriers of cassette tape and CD, the concept of printed journals or even journals in the conventional sense is not important any more. Actually, for some new journals, articles are not organized and published by issues and volumes, e.g., PLOS ONE, Scientific Reports, eLIFE and Peer J, etc.

It is necessary to make changes to the current research evaluation way rooted in the journal selection system. We should be aware of that journal evaluation is not equal to article evaluation, evaluating scientists, institutions and countries based on article evaluation is more reasonable than the current journal-based evaluation. In order to make better assessment of research performance and research excellence, we propose the idea of article level evaluation system and database. Using metrics data at different time periods of 46 articles in one issue, we make empirical test of the article level evaluation method.

Firstly, the basic function of this evaluation system is to assess the qualities of articles. Based on article level evaluation, it is also available to assess the research excellence of scientists, journals, institutions and countries. For example, how many articles tracked in phase 3 and 4 are published by one specific institution? What are the top institutions in one specific field? Secondly, both scholarly and societal impact of articles are taken into account. Thirdly, using the article usage data and online mention data, we can make evaluation of newly published papers. At different phases after publication, the comprehensive score of the paper is calculated with different weights of metrics, so the score and rank of a paper in different phases change.

To accomplish this, the biggest problem needs to be solved is the availability of metrics data. The citation data could be obtained from Web of Science, Scopus, Google Scholar, etc. The online attention data, e.g., social media, news reports, Mendeley readership is also available from various but certain data sources. However, for the article usage data, only part of academic publishers and journals provide usage data to public, including Nature Publishing Group, Science, PLOS, Taylor & Francis, ACM Digital Library, IEEE Xplore Digital Library, etc (Wang, Mao, Xu, & Zhang, 2013). For many others, e.g., Elsevier, Sage and Wiley, they may provide the metrics data of each article to some specific users and subscribers, but not free to public. If we want to evaluate all the papers whatever the publishers are, metrics data from publishers is indispensable.

With the movement from print to electronic publishing and the diversification of article-level-metrics, it is time to make change to the current research evaluation system. To better assess scientists' research and satisfy the evaluation needs in many situations, ranging from funding decisions to hiring tenure and promotion, we need to build an article-level-evaluation system. It is not an easy job, of course there are lots of issues need to be resolved, but article-level evaluation is the right way to make assessment of researches, we are moving on the right direction.

## Acknowledgements

The work was supported by the project of ''National Natural Science Foundation of China'' (61301227), and the project of "Growth Plan of Distinguished Young Scholar in Liaoning Province"(WJQ2014009).

## References

Alberts, B. (2013). Impact factor distortions. *Science, 340*(6134), 787-787.

Bordons, M., Fernández, M. T., & Gomez, I. (2002). Advantages and limitations in the use of impact factor measures for the assessment of research performance. *Scientometrics, 53*(2), 195-206.

Garfield, E. (2006). The history and meaning of the journal impact factor. *Jama, 295*(1), 90-93.

Opthof, T. (1997). Sense and nonsense about the impact factor. *Cardiovascular research, 33*(1), 1-7.

PLoS_Medicine_Editors. (2006). The impact factor game. *PLoS medicine, 3*(6), e291.

Priem, J., Taraborelli, D., Groth, P., & Neylon, C. (2010). Altmetrics: A manifesto. In (Vol. 2014). http://altmetrics.org/manifesto.

Saaty, T. L. (1980). *The analytic hierarchy process: planning, priority setting, resources allocation*. New York: McGraw.

Seglen, P. O. (1997). Why the impact factor of journals should not be used for evaluating research. *Bmj, 314*(7079), 497.

Wang, X., Liu, C., Fang, Z., & Mao, W. (2014). From Attention to Citation, What and How Does Altmetrics Work? *arXiv preprint arXiv:1409.4269*.

Wang, X., Mao, W., Xu, S., & Zhang, C. (2013). Usage History of Scientific Literature: Nature Metrics and Metrics of Nature Publications. *Scientometrics, 98*(3), 1923-1933.

Wang, X., Mao, W., Zhang, C., & Liu, Z. (2013). The diffusion of scientific literature in web. In *STI 2013 – 18th International Conference on Science and Technology Indicators* (pp. 415-426). Berlin.